# THERMAL RADIO EMISSION AS INCOHERENT INDUCED RADIATION

Fedor V.Prigara

*Institute of Microelectronics and Informatics, Russian Academy of Sciences, 21 Universitetskaya, 150007 Yaroslavl, Russia*

-----------------------------------------------------------------------------------------------------------

**The synchrotron interpretation of radio emission from various astrophysical objects[1] encounters the essential difficulties. These difficulties may be overcome in a new theory of thermal radio emission of non-uniform gas[2] which is based on a stimulated character of thermal radio emission following from the relations between Einstein's coefficients[3,4] for spontaneous and induced emission of radiation .**

Although thermal blackbody radiation did played a significant role in the origin of quantum theory[3], a further development of quantum mechanics pushed aside researches on thermal radiation . The experimental studies of thermal radio emission did not developed due to the low intensity of thermal radiation given by the Rayleigh-Jeans formula[1] .

As a result, in the early years of radio astronomy the properties of thermal radio emission of non-uniform gas were practically unknown . In particular it was not noticed that the relations between Einstein's coefficients for spontaneous and induced emission of radiation lead to an induced character of thermal radio emission[5] .

Owing to the induced origin of thermal radio emission radio waves with a wavelength $\lambda$ are emitted by a molecular resonator the size of which has an order of magnitude $l=1/(n\sigma)$, where *n* is the number density of particles (ions) , and $\sigma$ is the absorption cross-section .

Each molecular resonator emits coherently at the wavelength $\lambda=l$ , thermal radio emission of gas being incoherent sum of emission produced by individual emitters[5] .

The existence of maser sources associated with gas nebulae[6] is closely connected with the induced character of thermal radio emission .

Thermal radio emission of gas placed in magnetic field is polarized since an original spontaneous emission amplified by maser mechanism is polarized .

The above condition of emission implies that radio waves with a wavelength $\lambda$ are emitted by a gaseous layer with a definite number density of particles $n$ . The intensity of radiation is given by the Rayleigh-Jeans formula . Since an induced emission dominates, Kirchhoff's law for emissivity is not valid .

In the early years of radio astronomy it was believed that the spectral index $\alpha$ defined by the expression for the flux density $F_\nu \propto \nu^\alpha$ , where $\nu$ is the frequency of radiation , in the case of thermal radio emission cannot be negative . Meanwhile the disk model of extended radio source[2] together with the above condition of emission and some natural assumptions on the density and pressure distribution leads to the spectrum with $\alpha = -1$ .

It is worthwhile to note that the polarization of radiation and the existence of radio spectra with a negative spectral index were the main arguments in a favor of the synchrotron interpretation of radio emission from supernova remnants and radio galaxies in the early years of radio astronomy .

A further development of radio astronomy produced the essential discrepancies between the predictions of the synchrotron theory and observational data . According to the synchrotron theory a spectral index has a local origin and therefore the synchrotron theory is unable to explain the correlation between the radial distribution of brightness and the spectral index of emission from supernova remnants[7,8] .

The wavelength dependence of radio source size[2,9] also cannot be explained by the synchrotron theory . Other difficulties of the synchrotron theory are discussed in Ref.5 . Note that the synchrotron theory does not give the magnitude of the spectral index, which remains therefore an empirical parameter in this theory .

Thus , observational data support the thermal interpretation of radio emission from supernova remnants and extragalactic sources . The reason for the misinterpretation of observational data in the early years of radio astronomy is that the induced character of thermal radio emission was not taken into account . Thermal origin of emission from active galactic nuclei is supported also by the recent observations of the polarization of optical lines in the spectrum of radio galaxy coincident with the polarization of continuum[10] .

-------------------------------------------------------------------------------------------------------------